\documentclass[10pt]{article}

\usepackage{amsmath}
\usepackage{amssymb}

\usepackage{graphicx}

\usepackage{cite}

\usepackage{color}

\usepackage[labelfont=bf,labelsep=period,justification=raggedright]{caption}

\makeatletter
\renewcommand{\@biblabel}[1]{\quad#1.}
\makeatother

\date{}

\newcommand{\beq}{\begin{equation}}
\newcommand{\eeq}{\end{equation}}
\newcommand{\beqa}{\begin{eqnarray}}
\newcommand{\eeqa}{\end{eqnarray}}
\renewcommand{\a}{\alpha}
\renewcommand{\b}{\beta}
\newcommand{\cas}{\noindent $\bullet$ {\hskip 3pt}}
\newcommand{\e}{{\rm e}}
\newcommand{\eps}{\varepsilon}
\newcommand{\g}{\gamma}
\newcommand{\half}{{\textstyle{\frac{1}{2}}}}
\newcommand{\mud}{{\mu_{\rm d}}}
\newcommand{\mus}{{\mu_{\rm s}}}
\newcommand{\st}{{\rm st}}
\newcommand{\xid}{{\xi_{\rm d}}}
\newcommand{\xis}{{\xi_{\rm s}}}

\begin{document}

\begin{flushleft}
{\Large
\textbf{Synaptic metaplasticity underlies tetanic potentiation in {\it Lymnaea}: a novel paradigm}
}
\\
Anita Mehta$^{1,\ast}$,
Jean-Marc Luck$^{2}$,
Collin C. Luk$^{3}$,
Naweed I. Syed$^{3}$
\\
\bf{1} S. N. Bose National Centre for Basic Sciences, Calcutta, India
\\
\bf{2} Institut de Physique Th\'eorique, CEA Saclay and CNRS URA 2306, Gif-sur-Yvette, France
\\
\bf{3} Hotchkiss Brain Institute, Faculty of Medicine, University of Calgary, Alberta, Canada
\\
$\ast$ E-mail: anita@bose.res.in
\end{flushleft}

\section*{Abstract}

We present a mathematical mo\-del which explains
and interprets a novel form of short-term potentiation, which was
found to be use-, but not time-dependent, in experiments done on
{\it Lymnaea} neurons.
The high degree of potentiation
is explained using a model of synaptic metaplasticity, while the
use-dependence (which is critically reliant
on the presence of kinase in the experiment) is explained using a mo\-del of
a stochastic and bistable biological switch.


\section*{Introduction}
\label{intro}

All brain functions,
ranging from simple reflexes to complex motor patterns, learning and memory,
rely upon synaptic transmission between neurons through
specialized structures termed synapses.
These synaptic connections are, however, not static in nature; in fact,
they exhibit a high degree of synaptic plasticity,
enabling a network to generate behaviorally relevant and
functionally meaningful outputs.
These changes in synaptic strength can
either be short- or long-term, and may form the basis for both short- and
long-term memory, respectively.

Synaptic plasticity is neither restricted to any select group of neurons nor to a
particular species -- rather it is a universal trademark of all neurons that
have been investigated to date.
When a nervous system is unable to exhibit modulatory changes
associated with short- and long-term synaptic plasticity,
it is rendered dysfunctional.
Therefore, defining the mechanisms underlying synaptic plasticity is not only
pivotal for our understanding of how the brain functions but also for managing the
behavioral, learning, memory and cognitive defects that are met in clinical practice.
However, despite recent advances in our understanding of various
modes of neuronal communication, the cellular and molecular mechanisms
underlying synaptic plasticity remain poorly defined.
Moreover, the data generated from experimental studies has often been inadequate
to garner mathematical modeling predictions that may aid future research in
this area, or to provide insights into the mechanisms of synaptic plasticity.
This field could however benefit from a paradigm shift
where modeling approaches could be used to predict
elements of synaptic plasticity and to facilitate future research
in the area of metaplasticity.

In a recent study~\cite{luk}, two of the authors of this paper
observed a form of short-term potentiation
induced by tetanic stimulation, whose time-frame exceeded conventional forms
of short-term potentiation.
While the induction of potentiation was similar to previous forms
of short-term potentiation,
the time frame of the potentiated response was characteristic
of long-term potentiation ($\sim$ 5 hrs).

\begin{figure}[!ht]
\begin{center}
{\hskip 9pt}\includegraphics[angle=0,width=.5\linewidth]{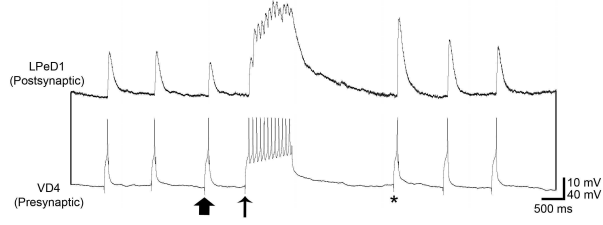}
\caption{\label{Figure1}
A representative electrophysiology trace showing the potentiation of the VD4/LPeD1
synapse following tetanic stimulation and its subsequent depotentiation.
A tetanic stimulation of presynaptic neuron VD4 (at thin arrow) results in
a compound excitatory postsynaptic potential (EPSP) in postsynaptic neuron LPeD1.
A subsequently triggered action potential in VD4 (asterisk) results in an EPSP
of greater amplitude than pre-tetanic stimulation (thick arrow).}
\end{center}
\end{figure}

More specifically, using well-defined, excitatory cholinergic
synapses between {\it Lymnaea} pre- and postsynaptic neurons, specifically
visceral dorsal 4 (VD4) and left pedal dorsal 1 (LPeD1-Excitatory), they provided
evidence for a novel form of short-term potentiation, which was use-, but not
time-dependent.
They found that following a tetanic stimulation ($\sim$ 10 Hz) in the presynaptic
neuron with a minimum of seven action potentials,
the synapse became potentiated, whereby a subsequent action
potential triggered in the presynaptic neuron resulted in an enhanced
postsynaptic potential (see Figure~\ref{Figure1}).
Further, if an inducing tetanic stimulation was activated
but a subsequent action potential was not triggered,
the synapse was shown to remain potentiated for as long as 5 hours.
However, once this action potential was triggered,
the authors found that the synaptic strength rapidly returned to baseline levels.
It was also shown that this form of synaptic plasticity relied on the
presynaptic neuron, and required pre- but not postsynaptic
Ca$^{2+}$/calmodulin dependent kinase II (CaMKII) activity.
Hence, this form of potentiation shares induction and de-potentiation
characteristics similar to other forms of short-term potentiation,
but exhibits a time-frame analogous to that of long-term potentiation.
The model of metaplastic synapses~\cite{meh} reviewed below allows us
to reproduce these long timescales, and forms the basis of our explanation
of the experimental results reported in~\cite{luk}.

In the model of metaplastic synapses, incoming signals are
stored as memories at progressively deeper levels of a synapse, leading to a
clear temporal separation between long- and short-term memory.
The upper levels are more vulnerable to `noise', i.e., the regular influx of
(usually irrelevant) information which is responsible for the phenomenon of forgetting;
only short-term memories can be stored here.
(For a review of noise in neural systems, see~\cite{fai}).
Deeper synaptic levels
are more protected from this noise, and thus able to retain the memory of applied
signals for a much longer time.
Drawing on these structural ideas first proposed by Fusi et al.~\cite{fus},
the model of~\cite{meh}
provides a theoretical framework for the {\it dynamics} of
signal propagation within the metaplastic synapse.
It suggests that
random signals are typically stored in the upper levels of the synapse for
relatively short times,
and then lost to noise: non-random signals, on the other hand, are stored
as long-term memories in the deepest synaptic levels and forgotten much more slowly.

The link between this theory and the experiment reported in~\cite{luk} relies
on the fact that the output signal in the latter was {\it amplified}
after a process of tetanic stimulation.
This suggested the following scenario:
the initial action potentials, interpreted as a non-random signal,
cumulatively built up a long-term
memory of the signal in the deepest synaptic levels.
The synapse dynamics were then frozen so that further discharge was prevented.
When a further action potential was applied, the synaptic dynamics restarted (`use'-dependence):
the release of the accumulated memory from the deepest levels of the synapse
constituted the observed enhancement of the output signal described in~\cite{luk}.
While this enhancement is plausibly accounted for by
the model of metaplastic synapses~\cite{meh}, the explanation of the
freezing of the synaptic dynamics and its subsequent use-dependence needs the introduction of a biological switch.
The stochastic and bistable switch presented in this paper meets this need, and models the role of kinase
(CaMKII) in the actual experiment~\cite{luk}.

In the following,
we first review the basics of the model of a metaplastic synapse~\cite{meh}.
We then provide a full theoretical framework for the explanation
of the experimental results, with an emphasis on the dynamics of the biological switch.
We close by discussing our results.

\section*{Results}

\subsection*{A model of a metaplastic synapse}

If synapses are highly plastic, signals are quickly stored: however,
high plasticity also means that more and more signals are stored, generating
enough `noise' so that `memories' of earlier signals soon become irretrievable.
Clearly, this is at variance with the fact that long-term memories
are ubiquitous in human experience;
it was to resolve this paradox that models of metaplastic synapses
were formulated~\cite{fus}.
The idea behind such models was that the introduction
of `hidden states' for a synapse would enable the delinking of
memory lifetimes from instantaneous signal response: while maintaining quick
learning, this mechanism would also be able to allow slow forgetting.
This was implemented
by the storage of memories at different `levels': the relaxation times for
the memories increased as a function of depth.
This hierarchy of time scales models
the phenomenon of metaplasticity~\cite{abr,fis}.

In~\cite{meh} these ideas were put into a new
framework, with the dynamics of signal processing playing a central role.
Also, two different internal synaptic structures were investigated:
the first (Model~I) was very similar to Fusi's original model~\cite{fus},
while the second (Model~II) had a different architecture.
In the following, we focus on the second model.
We start by reviewing its essential features.

The dynamics of the model are defined in Figure~\ref{II}.
At every discrete time step $t$, the synapse is subjected either to
a potentiating pulse (PP) (encoded as $\eps(t)=+1$)
or to a depressing pulse (DP) (encoded as $\eps(t)=-1$),
where $\eps(t)=\pm1$ is the instantaneous value
of the input signal at time $t$.
There are three outcomes of the application of a PP signal:

\cas
If the synapse is in its $-$ state at depth $n$,
it may climb one level $(n\to n-1)$ with probability $\a_n$.

\cas
If it is in its $-$ state at depth $n$,
it may alternatively cross over to the $+$ state {\it at the same level} with
probability $\b_n$.

\cas
If it is already in its $+$ state at depth $n$,
it may fall one level $(n\to n+1)$ with probability $\g_n$.

The {\it level-resolved output signal} of level $n$ at time $t$:
\beq
D_n(t)=Q_n(t)-P_n(t)
\label{dndef}
\eeq
and the {\it total output signal} at time $t$:
\beq
D(t)=\sum_{n\ge0}D_n(t)
\label{ddef}
\eeq
can be expressed in terms of the probabilities $P_n(t)$ (resp. $Q_n(t)$)
for the synapse to be in the $-$ state (resp. in the $+$ state)
at level $n=0,1,\dots$ at time $t=0,1,\dots$
The dynamical equations obeyed by the latter probabilities
are reviewed in the Methods section,
along with other details for the mathematically inclined reader.

Before any meaningful signal is applied,
the synapse is assumed to be in its default state.
The latter state is defined as
the stationary state reached by the synapse if subjected to a long random input signal.
It is described in detail in the Methods section
(see equations~(\ref{randef}) to~(\ref{xis})).
When a single potentiating pulse signal is applied at time $t=1$ (that is, $\eps(1)=+1$)
to the synapse in its default state,
the synapse will get polarized in response, and thus `learn' the signal.
Later on, under the influence of a random input signal for times $t\ge2$,
it will `forget' the PP signal, and return to its default state.
Figure~\ref{ltpred} shows plots of the reduced output signal $D(t)/D(1)$
against time $t$ for several values of the control parameter $\b$.
All subsequent figures refer to the parameter values $\b=0.2$, $\g=0.5$,
and $\xis=\xid=5$ (see Methods section).
From here on, we will refer to times where the synapse is subjected to a
significant signal ($\eps(t)=\pm1$) as {\it learning phases},
and to times where the synapse is subjected to random input
($\eps(t)=0$) as {\it forgetting phases}.

The late stages of the forgetting process are characterized by a universal power-law decay of the output signal:
\beq
D(t)\sim t^{-\theta}.
\label{dtheta}
\eeq
This is known as {\it power-law forgetting}~\cite{wix1,wix2,bro}.
The forgetting exponent
\beq
\theta=1+\frac{\xid}{\xis}
\label{theta}
\eeq
is always larger than unity and depends on the ratio of the dynamical and static lengths
$\xid$ and $\xis$.
If the synapse were finite rather than infinite, and consisted of $N$ levels,
the power-law decay (\ref{dtheta}) would be exponentially cut off at a time
\beq
\tau_N\sim\exp(N/\xid)
\eeq
which grows exponentially fast with the ratio
of the number $N$ of levels to the dynamical length $\xid$.

We now describe the effect of a sustained input of potentiating pulses
lasting for $T$ consecutive time steps ($\eps(t)=+1$ for $1\le t\le T$)
on the model synapse: in the following,
this will be referred to as a long-term potentiating (LTP) signal.
The synapse is again assumed to be initially in its default state.
The learning and forgetting processes are qualitatively similar
to the PP case described above,
while novel qualitative features emerge
when the duration of the signal is long enough ($\b T\gg1$).
In this regime, the synapse gets almost totally polarized
under the persistent action of the input signal.
This saturation phenomenon is illustrated in Figure~\ref{sted},
which shows the output signal $D(t)$
for several durations $T$ of the LTP signal.

The synapse slowly builds up a long-term memory in the presence
of a long enough LTP signal, as the memorized signal moves to deeper
and deeper levels.
At the end of the learning phase ($t=T$),
the polarisation profile will have the form of a sharply peaked traveling wave,
around a typical depth which grows according to the logarithmic law
\beq
n(T)\approx\xid\ln\g T.
\label{nT}
\eeq
The total output signal then decays according to the universal power law (\ref{dtheta}),
irrespective of the duration of the learning phase, driving home the universality
of power-law forgetting.

The above model provides a mechanism for the long-term memory storage
due to use dependence in the {\it Lymnaea} synapse examined in~\cite{luk}.
With the tetanic stimulation acting as an LTP signal, this model shows that the synapse becomes fully polarized,
the memory of the stimulating pulses being `stored' in a deep level.
However, another concept is needed to model the subsequent freezing of the synaptic
dynamics:
a major clue is provided by the fact that it is suggested in~\cite{luk}
that the activation of CaMKII in the presynaptic cell acts like a 'molecular switch'.
Accordingly, we present a theoretical model of
a {\it stochastic and bistable} switch in the following section.

\subsection*{The effect of use-dependent synaptic potentiation:\\
coupling to a stochastic and bistable switch}

In this section, we present a model of a biological switch
to describe the role of CaMKII in the experiment reported in~\cite{luk}.
The switch can exist in two states, `on' or `off', which we label by
the binary variable $\sigma(t)=1$ or 0.
Since natural processes are usually
stochastic rather than deterministic, we incorporate this by postulating that
the switch is on with probability
$\Pi(t)$, and off with the complementary probability $1-\Pi(t)$.
Thus:
\beq
\sigma(t)=\left\{
\begin{matrix}
1\hfill&\hbox{with probability\ }\Pi(t),\hfill\cr
0\hfill&\hbox{with probability\ }1-\Pi(t).
\end{matrix}
\right.
\label{etadef}
\eeq

The main effect of this switch is to freeze the synaptic dynamics
after adequate potentiation: we accordingly refer to the probability $\Pi(t)$ as the {\it freezing probability}.
Thus:

\cas
If the switch is {\it off}
($\sigma(t)=0$), the synapse evolves as usual.
This occurs with probability $1-\Pi(t)$.

\cas
If the switch is {\it on} ($\sigma(t)=1$), the forgetting process (`discharge') is frozen.
This occurs with probability $\Pi(t)$.

More precisely,
the synapse learns via~(\ref{II+})-(\ref{II-})
and forgets via~(\ref{meanII}) when the switch is off.
When the switch is on, the synapse
still learns via~(\ref{II+})-(\ref{II-}),
but it does not forget at all.

In the experiment, a minimum of seven action potentials is needed
to activate the switch and freeze the dynamics: this suggests
that the switch somehow responds to the saturation of the synaptic capacity,
so that freezing {\it never} sets in for less than this number of action potentials.
Also,
a further tetanic pulse after a period of quietude is needed to restart the synaptic dynamics.
We design the evolution of the freezing probability $\Pi(t)$
accordingly:

\cas
If the synapse is within a learning phase
($\eps(t)=\pm1$ and $\eps(t-1)=\pm1$),
the freezing probability evolves according to the quadratic rule
\beq
\Pi(t)=1-c(1-\Pi(t-1))^2.
\label{pidyn}
\eeq
This rule ensures that the freezing probability increases with the number of
action potentials applied,
saturating quickly to $\Pi=1$ and freezing the synaptic dynamics after a
threshold number of these is reached.
It is desirable that the fixed point of $\Pi=1$ is superstable (see below),
and the quadratic law~(\ref{pidyn}) is the simplest non-linear law which ensures this.

\cas
If the synapse is in a forgetting phase ($\eps(t)=0$),
the freezing probability is itself frozen to its value inherited from the past:
\beq
\Pi(t)=\Pi(t-1).
\eeq
This rule ensures that once frozen, the dynamics stay frozen
and that the synapse stays potentiated, until a further action potential is applied.

\cas
If the synapse is at the first step of a learning phase
($\eps(t)=\pm1$ but $\eps(t-1)=0$), the freezing probability is instantly reset to
\beq
\Pi(t)=0.
\eeq
This ensures that the synaptic dynamics restart, as
soon as an action potential is applied for the first time following a period of forgetting.

Note that there is a `soft' threshold in the experiment
for the switch to kick in, in that a {\it minimum} of seven
action potentials is required; as reported in~\cite{luk},
this phenomenon was usually seen to occur across a range of 7 to 14 action potentials.
The freezing probability which models the switch
dynamics needs to incorporate this soft threshold, for which
superstability of the fixed point $\Pi=1$ is desirable.
This ensures that at the end of a sustained LTP signal of duration~$T$,
$\Pi(T)$ converges very rapidly to unity (more rapidly than exponentially),
as soon as $T$ exceeds a characteristic time $T_0$,
defined operationally by $\Pi(T_0)=\half$.
The value of the characteristic time fixes the parameter $c$.
We have then
\beq
\Pi(T)=1-\exp\left(-\frac{2^{T-1}-1}{2^{T_0-1}-1}\ln 2\right).
\eeq
Figure~\ref{Pi} shows a plot of the freezing probability $\Pi(T)$
at the end of an LTP signal of duration $T$
for several values of the characteristic time $T_0$ in the realistic range
of $5\le T_0\le 9$.
In practice, as soon as $\Pi(T)$ is appreciably large,
the switch kicks in stochastically -- and this can occur,
as shown in Figure~\ref{Pi}, over a range of signal durations.

We now commence a global interpretation of the experiment.
The synapse is assumed to be initially in its default state, with $\Pi(0)=0$.
It is then subjected to a sustained LTP signal of duration $T_1$
(i.e., the application of $T_1$ action potentials),
and to a single action potential at a much later time ($T_2\gg T_1$).
The synapse is subjected to a random input
at all the other instants of time
($\eps(t)=+1$ for $1\le t\le T_1$ and for $t=T_2$, else $\eps(t)=0$).

In the regime where the number of action potentials $T_1$ of the initial signal
is larger than the characteristic time $T_0$ of the switch,
the freezing probability $\Pi(T_1)$ at the end of the LTP period is very high,
i.e., very close to unity (see Figure~\ref{Pi}).
During this learning phase,
the output signal $D(t)$ grows progressively from $D(0)=0$
to a large value $D(T_1)$.
The high value of $\Pi(T_1)$ at the end of this phase
typically freezes the synaptic dynamics,
ensuring that this enhanced output signal is not discharged.
When the next action potential is applied at time $T_2$, the switch is turned off,
and the synapse then relaxes via
the full discharge of the stored, enhanced output signal.

We now compare theory with experiment.
Figure~\ref{FigureX} shows a comparison between
our theoretical predictions (upper panel)
with sharp-electrode electrophysiology recordings
of a VD4/LPeD1 synaptic pair (two lower panels).
The theoretical prediction is meant to describe the average over many ensembles,
while the experimental data are assumed
to be typical. Since the experimental averages
are well-behaved rather than subject to large fluctuations~\cite{luk},
a typical experimental output is representative of its average.
On the theoretical side, the time unit for the discrete updates
is consistently chosen to be the time interval between two successive applied APs,
i.e., 100 ms.
The black theoretical curve corresponds to
3 APs triggered during tetanic stimulation,
which are insufficient to result in potentiation of a subsequent
excitatory postsynaptic potential (EPSP) in the LPeD1 neuron
($T_1=3\ll T_0$, and so the switch remains off).
The red theoretical curve corresponds to 11 APs,
resulting in a potentiated response
($T_1=11\gg T_0$, so the switch is turned on and the synapse is frozen).
The predicted pattern of peak heights (symbols)
is rather robust, i.e., insensitive to model parameters,
and provides a good overall description of the experimental recordings.

The last part of this section reflects the `bistability' of the switch,
i.e., its stability in one of two states.
If the duration $T_1$ of the initial LTP signal
is comparable to the characteristic time $T_0$,
the synapse may exhibit two types of temporal behavior.
It may behave as above for $T_1=11$,
i.e., stay frozen until the subsequent action potential unfreezes it at time $T_2$.
This occurs with probability $\Pi(T_1)$ (see Figure~\ref{Pi}).
The synapse may also remain unfrozen, so that the output signal decays almost entirely
before the subsequent action potential sets in.
This occurs with the complementary probability $1-\Pi(T_1)$.
This bistable behavior is illustrated in Figure~\ref{switch}.
We see from this figure that the difference between $T_1$ and $T_0$
provides a tunable parameter which determines how frequently
the phenomenon described in the experiment in~\cite{luk} is observed:
in the upper panel the parameters are such that it is almost always observed,
whereas in the lower panel it makes a much more random appearance.
This parametrisation is useful, since it can form the basis of future experiments.
If we assume that the characteristic time $T_0$ is, say, inversely proportional
to the concentration of kinase, this suggests that a threshold concentration
can be identified for which a given number of action potentials will cause the short-term
potentiation of the synapse.

\section*{Discussion}

Our theoretical model of a biological switch coupled to a metaplastic synapse is able
to provide an interesting framework for the interpretation of the experimental results
of~\cite{luk}.
This novel form of synaptic potentiation exhibits characteristics
of short-term potentiation,
but also exhibits characteristics of long-term potentiation based on its time frame.
While the enhanced output signal observed in the latter is
attributable to long-term potentiation of the synapse via tetanic stimulation,
the use-dependence of the synapse is explainable by the biological switch.

We start with the depotentiated state of the synapse,
where, experimentally, single action potentials triggered in VD4
elicit EPSP in LPeD1 of a non-potentiated amplitude.
The application of repeated action potentials in the pre-synaptic cell however
builds up long-term memory in the synapse, i.e., a potentiated EPSP in LPeD1.
Meanwhile, during the tetanic stimulation, the freezing probability of the molecular
switch (CaMKII) evolves according to~(\ref{pidyn})
in response to the growing saturation of the synapse.
If the number of action potentials is large enough,
the switch is activated and synaptic dynamics are fully frozen.
Synaptic dynamics are only restarted when the switch is turned off: this happens when the synapse is next `used', and
leads to the observed rapid decay back to a non-potentiated amplitude in LPeD1.
Thus, the critical dependence of the onset of the decay on synapse {\it use}, rather than the {\it time} elapsed after priming,
is explained by the model presented in this paper.

It is useful to discuss some of the key assumptions made here
before concluding.
The level of description of our model is sufficiently
abstract that it does not claim to replicate intricate biological detail; its
description is limited to probing the unusually long times of use-dependent synaptic
potentiation in the Lymnaea synapse.
From this point of view, the use of
a model of long-term memory in metaplastic synapses is perfectly appropriate.
This model embodies the idea that forgetting takes place via the exposure of the
multi-level synapse to noise, from which the lowest synaptic levels are protected.
This noise
could refer to fluctuations in neural/synaptic activity or to background
signals; what is clear is that it is ubiquitous and unavoidable.
The model of a biological switch which we introduce
depends both on the presynaptic cell and on the state of the synapse: it is turned
on when the synapse is saturated in response to a series of action potentials, leading
to the freezing of synaptic dynamics.
It is turned off when the synapse is next `used', i.e
an activation potential is applied after an inert period.
An enhanced output signal
results, because of the discharge of accumulated memory at the synapse.
The turning
on and off of the switch is stochastic, rather than deterministic, given the nature
of most natural processes: and it is bistable, since it can lead to two kinds of temporal
behaviour at the synapse (with different probabilities, of course).
Finally, since the experimental recordings do not show strong fluctuations~\cite{luk},
the output of a typical experiment
can be meaningfully compared with the average output signal given by theory.

While the intact brain is composed of a significantly greater number
of synapses with far more intricate connectivity patterns,
the current model has taken a reductionist approach
to understanding synaptic plasticity at the level of a single synapse.
While previous forms of synaptic potentiation have been modeled,
this novel form of use-dependent synaptic potentiation has not.
Therefore, this study may not only be an important step
in developing more complex models composed of multiple synapses,
but also be important in guiding
further research in understanding neuronal network function in intact {\it Lymnaea} brain.

\section*{Methods}

\begin{figure}[!ht]
\begin{center}
\includegraphics[angle=0,width=.15\linewidth]{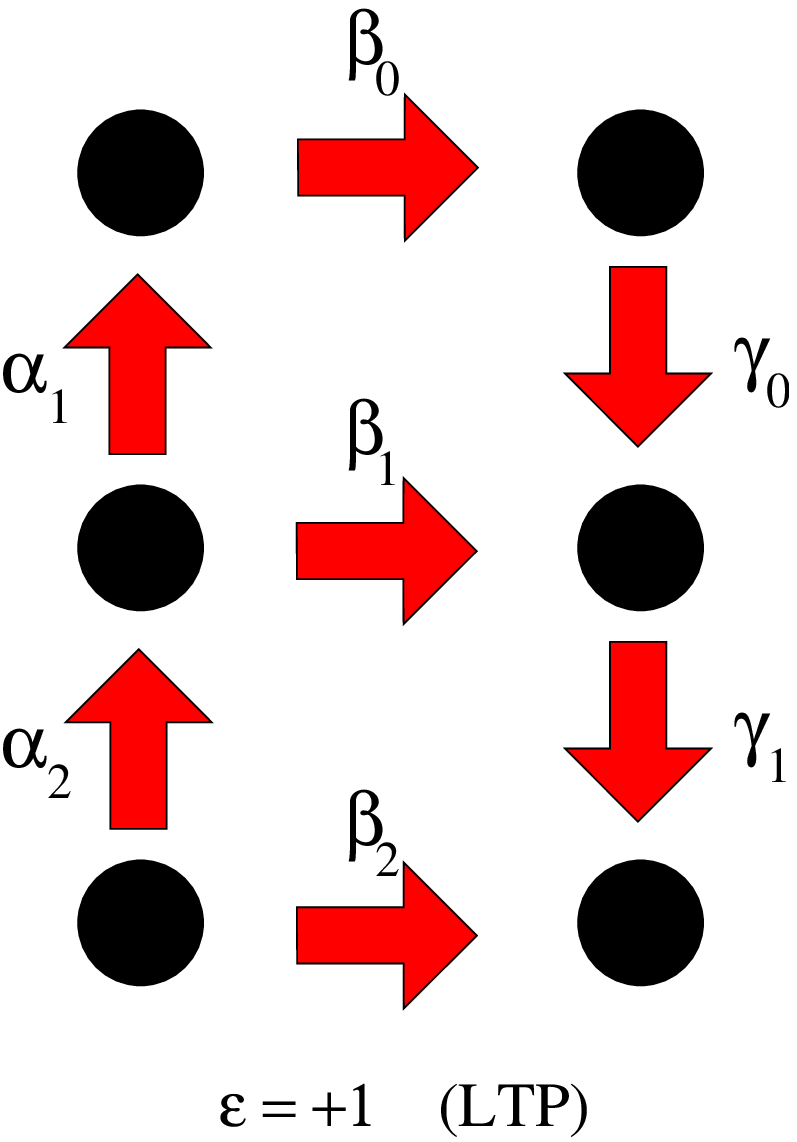}
{\hskip 20pt}
\includegraphics[angle=0,width=.15\linewidth]{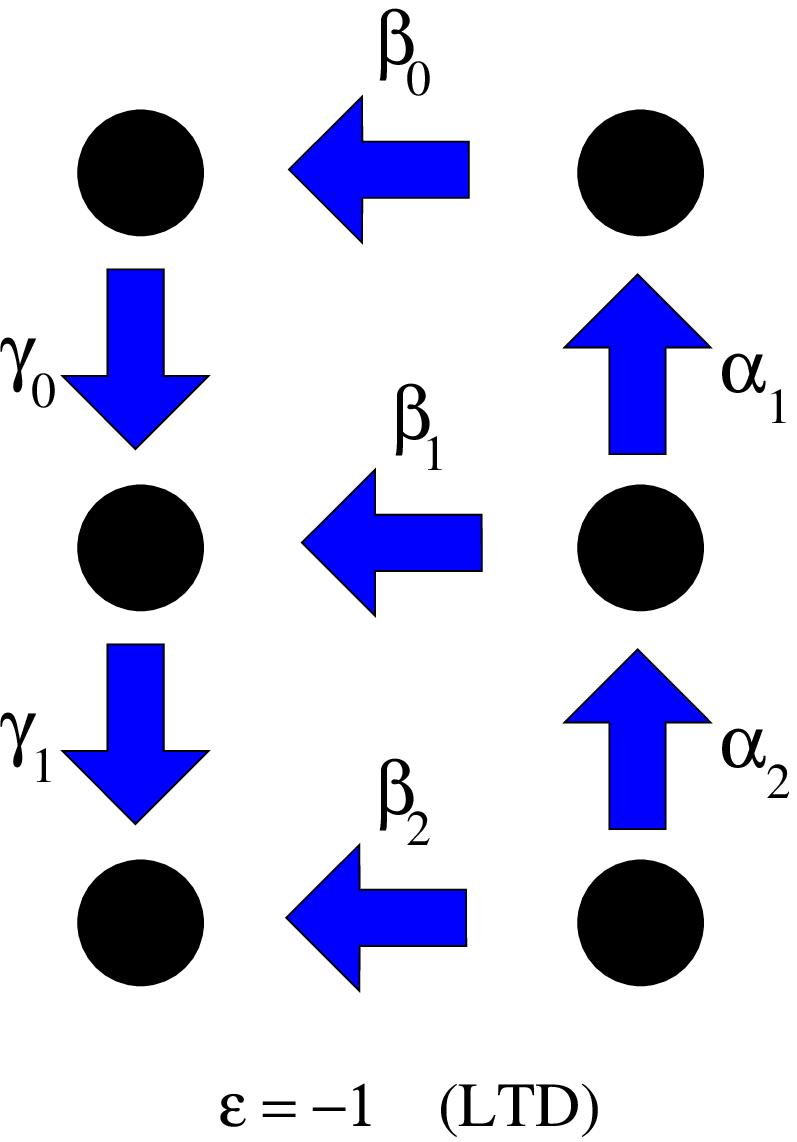}
\end{center}
\caption{
Schematic representation of the model of the internal synaptic structure.
Arrows denote possible transitions in the presence of a potentiating pulse
(PP, $\eps=+1$, left panel) and of a depressing pulse (DP, $\eps=-1$, right panel).
Corresponding transition probabilities are indicated.
In each panel, the left (resp. right) column corresponds to the $-$ (resp. $+$) state.
The model studied in this work is actually infinitely deep.
After~\cite{meh}.}
\label{II}
\end{figure}

\begin{figure}[!ht]
\begin{center}
\includegraphics[angle=-90,width=.3\linewidth]{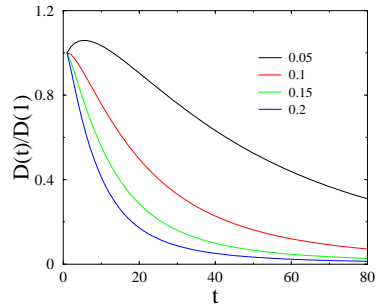}
\end{center}
\caption{
Plot of the reduced output signal $D(t)/D(1)$ after a single PP input signal,
against time $t$, for several $\b$.
After~\cite{meh}.}
\label{ltpred}
\end{figure}

\begin{figure}[!ht]
\begin{center}
\includegraphics[angle=-90,width=.3\linewidth]{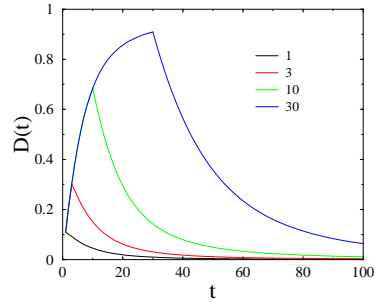}
\end{center}
\caption{
Plot of the output signal $D(t)$
against time $t$, for several durations $T$ of the LTP signal.
After~\cite{meh}.}
\label{sted}
\end{figure}

\begin{figure}[!ht]
\begin{center}
\includegraphics[angle=-90,width=.3\linewidth]{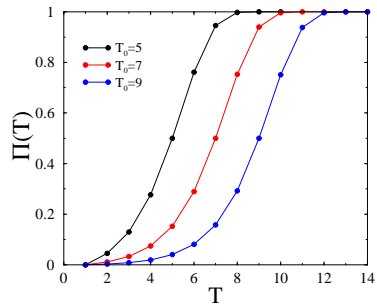}
\end{center}
\caption{
Plot of the freezing probability $\Pi(T)$
at the end of an LTP signal, against its duration $T$,
for three values of the characteristic time $T_0$.
The threshold for freezing is least for the black curve ($T_0=5$),
and most for the blue ($T_0=9$): while, say, 8 action
potentials will definitely cause the onset of freezing for the black curve
($\Pi(8)=0.997$), it will only rarely do so for the blue one ($\Pi(8)=0.292$).}
\label{Pi}
\end{figure}

\begin{figure}[!ht]
\begin{center}
\includegraphics[angle=-90,width=.5\linewidth]{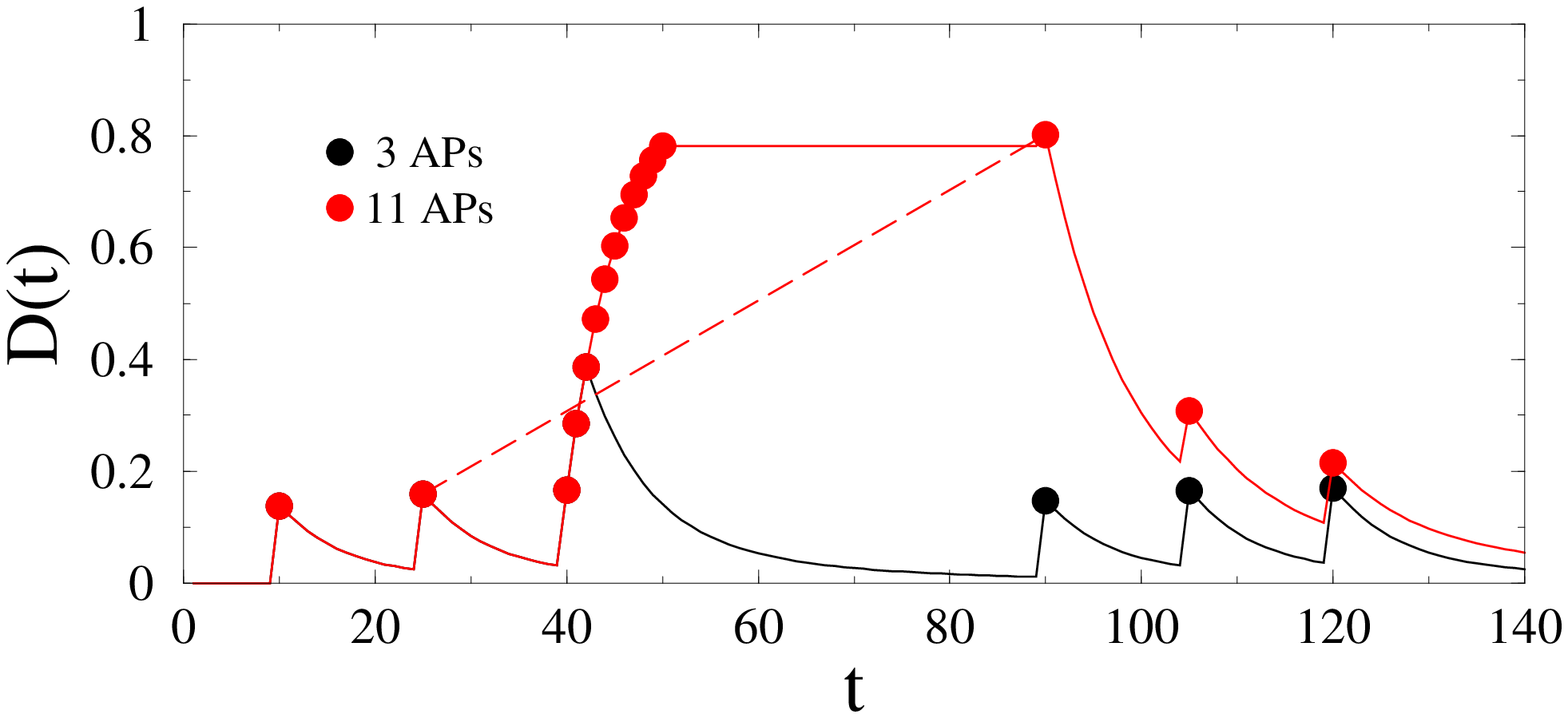}

{\hskip 8pt}\includegraphics[angle=0,width=.54\linewidth]{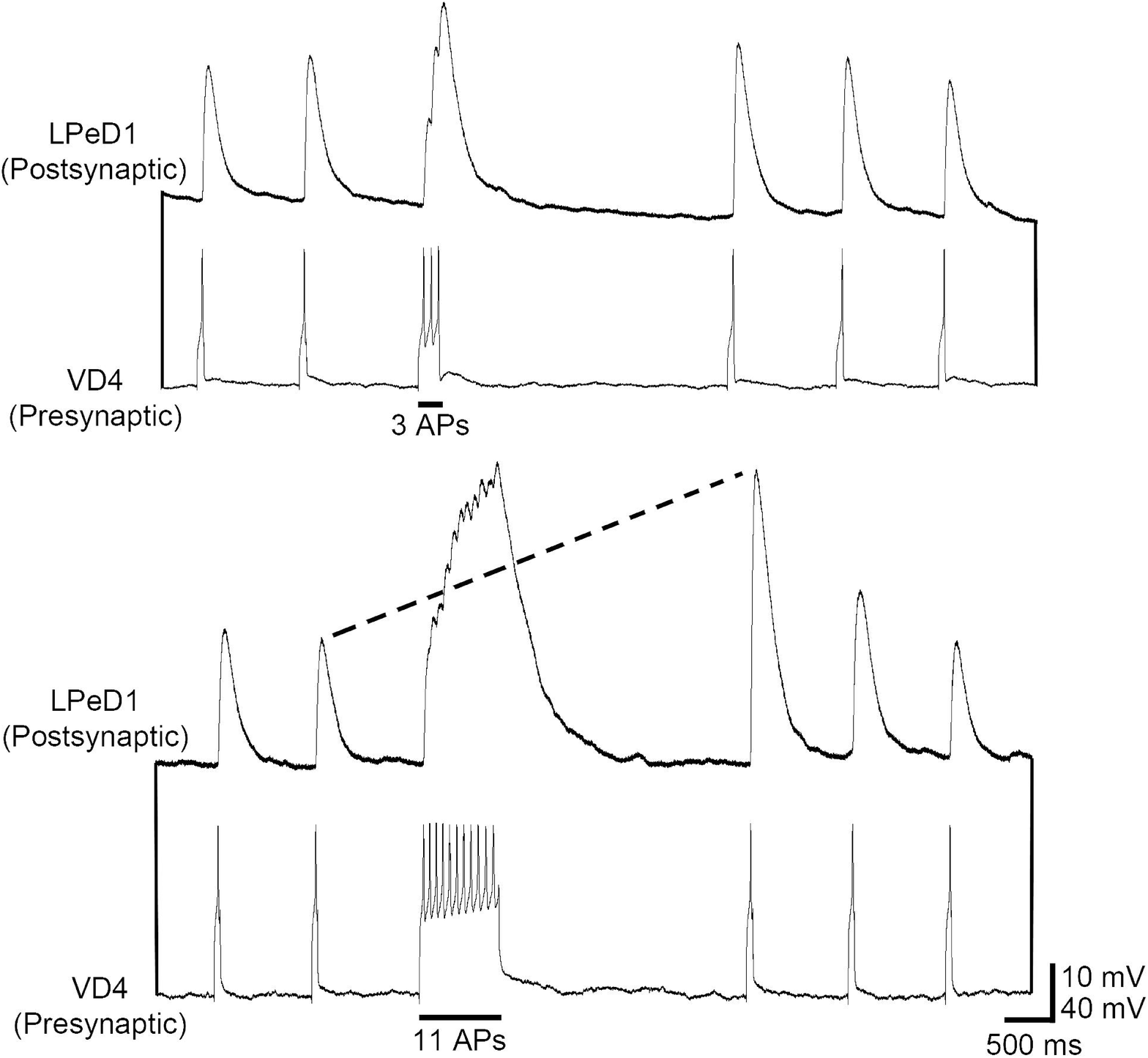}
\end{center}
\caption{
An integrative figure showing the good qualitative agreement of theory and experiment.
 The two lower panels show
 sharp-electrode electrophysiology recordings of a VD4/LPeD1 synaptic pair.
During the induction phase, three action potentials were triggered at  $\sim$ 10 Hz.
The synapse was allowed to remain quiescent for  $\sim$ 5 s
and when a subsequent action potential was triggered,
the amplitude of the postsynaptic potential was similar to pre-tetanic stimulation.
However, when eleven action potentials were triggered at  $\sim$ 10 Hz,
a potentiated response was observed
after the same quiescent period of  $\sim$ 5 s after
 stimulation. The upper panel shows the predictions of the theoretical model, also for
 three and eleven action potentials.}
\label{FigureX}
\end{figure}

\begin{figure}[!ht]
\begin{center}
\includegraphics[angle=-90,width=.3\linewidth]{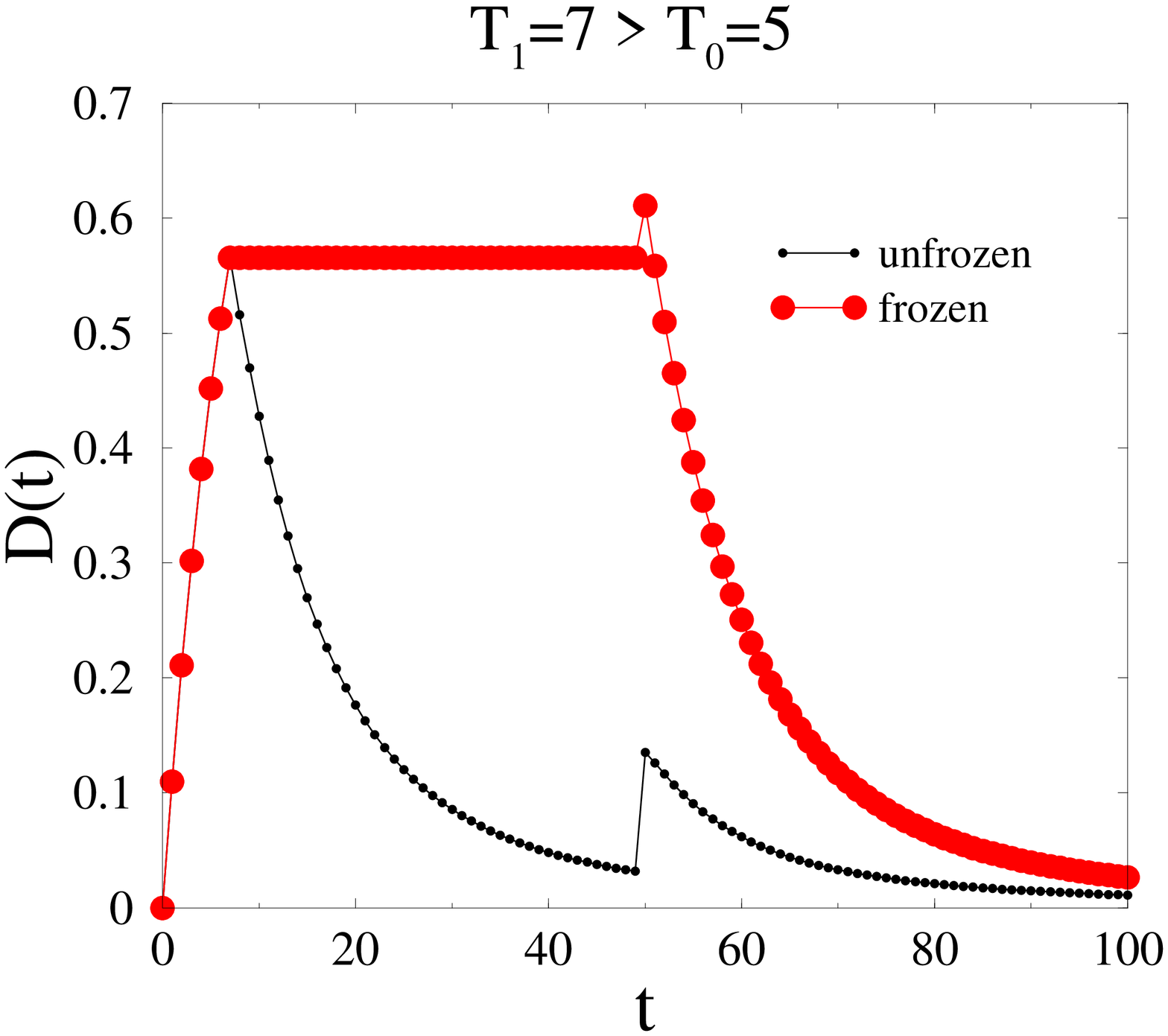}
{\hskip 10pt}
\includegraphics[angle=-90,width=.3\linewidth]{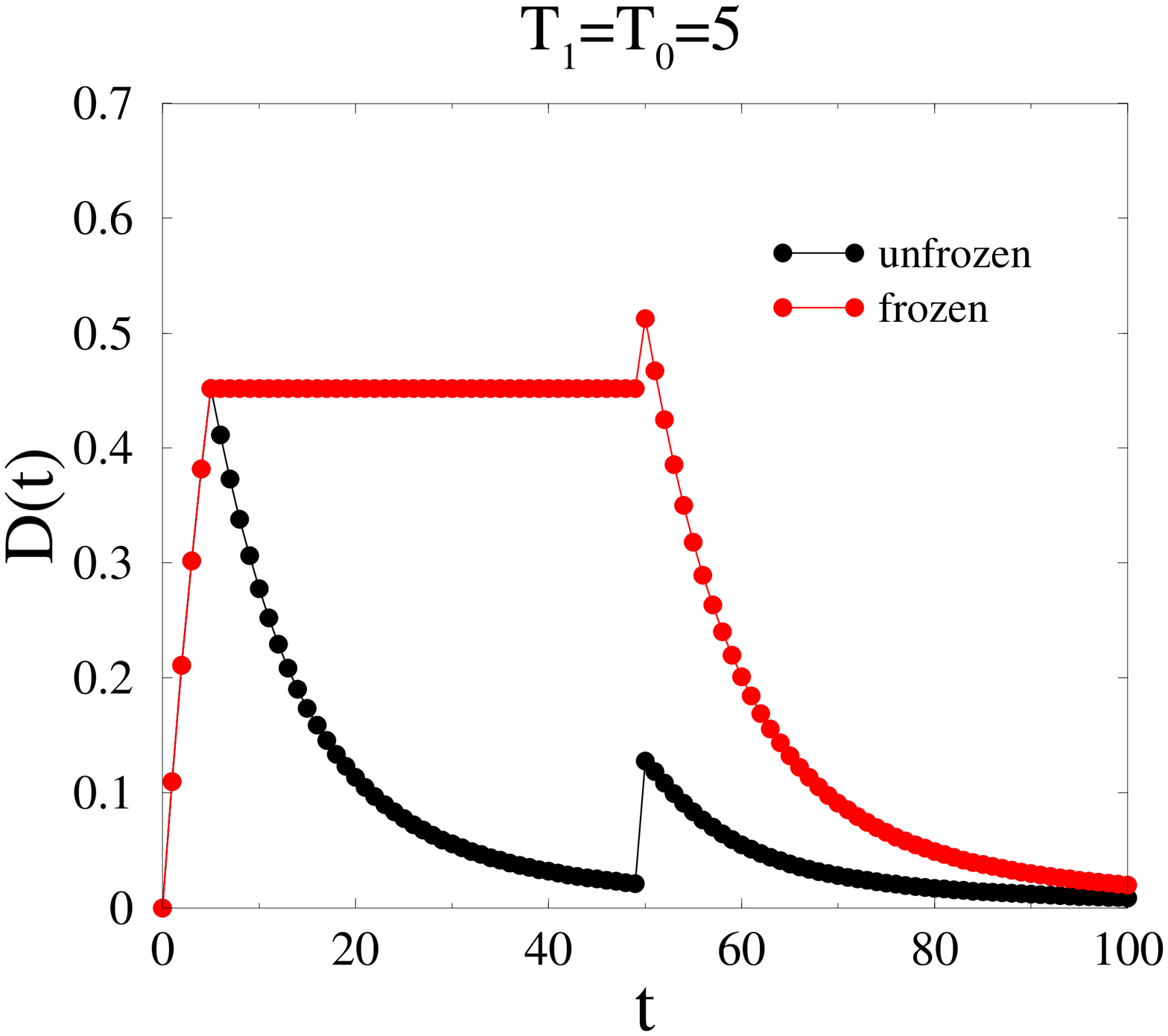}
\end{center}
\caption{
Plot of the two possible kinds of output signals $D(t)$
generated by the protocol described in the text, against time $t$
($T_0=5$, $T_2=50$).
Symbol sizes are proportional to the probabilities of each kind of behavior,
i.e., $\Pi(T_1)$ for the frozen one and $1-\Pi(T_1)$ for the unfrozen one.
Left: $T_1=7$ is larger than $T_0=5$, so that $\Pi(T_1)=0.946$ is very high.
Right: $T_1=T_0=5$, so that $\Pi(T_1)=\half$.}
\label{switch}
\end{figure}

The basic quantities used to describe the state of the synapse
are the probabilities $P_n(t)$ (resp. $Q_n(t)$)
for the synapse to be in the $-$ state (resp. in the $+$ state)
at level $n=0,1,\dots$ at time $t=0,1,\dots$
These probabilities obey the following coupled linear equations,
whose form is characteristic of Markov chains~\cite{van}:

\cas
$\eps(t+1)=+1$ (see Figure~\ref{II}, left):
\beqa
&&P_n(t+1)=(1-\a_n-\b_n)P_n(t)+\a_{n+1}P_{n+1}(t),\cr
&&Q_n(t+1)=(1-\g_n)Q_n(t)+\g_{n-1}Q_{n-1}(t)+\b_nP_n(t).
\label{II+}
\eeqa

\cas
$\eps(t+1)=-1$ (see Figure~\ref{II}, right):
\beqa
&&P_n(t+1)=(1-\g_n)P_n(t)+\g_{n-1}P_{n-1}(t)+\b_nQ_n(t),\cr
&&Q_n(t+1)=(1-\a_n-\b_n)Q_n(t)+\a_{n+1}Q_{n+1}(t).
\label{II-}
\eeqa

The transition probabilities of the model
are assumed to decay exponentially with level depth $n$:
\beq
\a_n=\a\e^{-(n-1)\mud},\quad
\b_n=\b\e^{-n\mud},\quad
\g_n=\g\e^{-n\mud},
\label{rates}
\eeq
where the {\it dynamical length}
\beq
\xid=\frac{1}{\mud}
\eeq
measures the number of fast levels at the top of the synapse.

The {\it default state} of the synapse is defined as
its stationary state in the presence of a random input signal,
defined by choosing at each time step
\beq
\eps(t)=\left\{
\begin{matrix}
+1\hfill&\hbox{with probability\ }\half,\cr
-1\hfill&\hbox{with probability\ }\half.
\end{matrix}
\right.
\label{randef}
\eeq

The dynamics of the synapse subjected to such a random input,
referred to as a `white-noise' random input in~\cite{meh},
is defined by averaging the linear dynamical equations
(\ref{II+})-(\ref{II-})
over both instances of $\eps(t)$ at each time step.
The resulting average dynamics is formally labeled as $\eps(t)=0$.
It has a simpler expression in terms of the sums and differences
\beq
S_n(t)=P_n(t)+Q_n(t),\quad
D_n(t)=P_n(t)-Q_n(t),
\eeq
namely
\beqa
&&S_n(t+1)=S_n(t)+\half
\bigl(\g_{n-1}S_{n-1}(t)-(\a_n+\g_n)S_n(t)+\a_{n+1}S_{n+1}(t)\bigr),\cr
&&D_n(t+1)=D_n(t)+\half
\bigl(\g_{n-1}D_{n-1}(t)-(\a_n+2\b_n+\g_n)D_n(t)+\a_{n+1}D_{n+1}(t)\bigr).
\label{meanII}
\eeqa

The default state of the synapse,
defined as the stationary state of the above average dynamics,
is characterized by the probabilities
\beq
P_n^\st=Q_n^\st=\half(1-\e^{-\mus})\e^{-n\mus},
\label{pqstat}
\eeq
with
\beq
\mus=\ln\frac{\a}{\g}.
\eeq
The default state is appropriately featureless and unpolarized,
as it should be for a symmetric synapse.
The corresponding {\it static length}
\beq
\xis=\frac{1}{\mus}
\label{xis}
\eeq
gives a measure of the effective number of occupied levels in the default state.

\section*{Acknowledgments}

A. M. thanks the Institut de Physique Th\'eorique,
where much of this work was carried out,
for its customary gracious hospitality during her visits.

\bibliography{plos}

\section*{Figure Legends}

\end{document}